\documentclass[10pt]{article}
\usepackage{graphicx}

\def\Title#1{\begin{center} {\Large #1 } \end{center}}
\def\Author#1{\begin{center}{ \sc #1} \end{center}}
\def\Address#1{\begin{center}{ \it #1} \end{center}}

\newcommand\pubblock{\rightline{\begin{tabular}{l} Proceedings of the Second Annual LHCP\\ \pubnumber\\
         \pubdate  \end{tabular}}}

\newenvironment{Abstract}{\begin{quotation} \begin{center} 
             \large ABSTRACT \end{center}\bigskip 
      \begin{center}\begin{large}}{\end{large}\end{center} \end{quotation}}

\newenvironment{Presented}{\begin{quotation} \begin{center} 
             PRESENTED AT\end{center}\bigskip 
      \begin{center}\begin{large}}{\end{large}\end{center} \end{quotation}}

\def\beq{\begin{equation}}
\def\eeq#1{\label{#1}\end{equation}}
\def\eeqn{\end{equation}}

\def\beqa{\begin{eqnarray}}
\def\eeqa#1{\label{#1}\end{eqnarray}}
\def\eeqan{\end{eqnarray}}

\let\bar=\overbar

\def\Dslash{\not{\hbox{\kern-4pt $D$}}}
\def\dslash{\not{\hbox{\kern-2pt $\del$}}}

\def\msb{{\bar{\ssstyle M \kern -1pt S}}}

\usepackage{color}

\newcommand\pubnumber{CMS CR-2014/186}

\newcommand\pubdate{\today}

\def\affiliation{
On behalf of the CMS Collaboration, \\
Department of Physics \\
INFN and University of Florence,
Italy}

\begin{document}

% large size for the first page
\large
\begin{titlepage}
\pubblock

%% Change the title, name, abstract
%% Title 
\vfill
\Title{Performance of the online track reconstruction and impact on hadronic triggers at the CMS High Level Trigger}
%\Title{Performance of hadronic triggers at the CMS High Level Trigger}
\vfill

%  if you need to add the support use this, fill the \support definition above. 
%   \Author{ FIRSTNAME LASTNAME \support }
\Author{Valentina Gori}
\Address{\affiliation}
\vfill
\begin{Abstract}
% HERE IS YOUR ABSTRACT.  REPLACE THE TEXT. The second annual Large Hadron Collider Physics (LHCP) conference will be held at Columbia University in New York City from June 2-7, 
% 2014. The conference is jointly hosted by Brookhaven National Laboratory and Columbia University. This conference is a result of a recent merger of two international conferences: 
% "Physics at Large Hadron Collider" and "Hadron Collider Physics Symposium".
The trigger systems of the LHC detectors play a crucial role in determining the physics capabilities of the experiments.
A reduction of several orders of magnitude of the event rate is needed to reach values compatible with the detector readout, offline storage and analysis capabilities.
The CMS experiment has been designed with a two-level trigger system: the Level 1 (L1) Trigger, implemented on custom-designed electronics, and the High Level Trigger (HLT),
a streamlined version of the CMS reconstruction and analysis software running on a computer farm. The software-base HLT requires a trade-off between the complexity of the
algorithms, the sustainable output rate, and the selection efficiency. This is going to be even more challenging during Run II, with a higher centre-of-mass energy,
a higher instantaneous luminosity and pileup, and the impact of out-of-time pileup due to the 25 ns bunch spacing. The online algorithms need to be optimised for such a
complex environment in order to keep the output rate under control without impacting the physics efficiency of the online selection. Tracking, for instance, will play
an even more important role in the event reconstruction. In this poster we will present the performance of the online track and vertex reconstruction algorithms, and their
impact on the hadronic triggers that make use of b-tagging and of jets reconstructed with the Particle Flow technique. 
We will show the impact of these triggers on physics performance of the experiment, and the latest plans for improvements in view of the Run II data taking in 2015.
\end{Abstract}
\vfill

% DO NOT CHANGE 
\begin{Presented}
The Second Annual Conference\\
 on Large Hadron Collider Physics \\
Columbia University, New York, U.S.A \\ 
June 2-7, 2014
\end{Presented}
\vfill
\end{titlepage}
\def\thefootnote{\fnsymbol{footnote}}
\setcounter{footnote}{0}
%

% normal size for the rest
\normalsize 

%% Your paper should be entered below. 

\section{The CMS trigger}
The collision rate at the Large Hadron Collider (LHC) is heavily dominated by large cross section QCD processes, which are not of prime interest for the physics program
of the CMS experiment. The processes relevant for new physics usually occur at a rate smaller than 10~Hz.
Since it is not possible to register all the events and to select them later on, because of a limited bandwith, it becomes mandatory to use a trigger system
in order to select events according to physics-driven choices.
The CMS experiment features a two-level trigger architecture.
The first level (L1), hardware, operates a first selection of the events to be kept, using muon chambers and calorimeter information.
The maximum output rate from L1 is about 100~kHz~\cite{tridas}; this upper limit is given by the CMS data acquisition electronics.
The second level, called \textit{High Level Trigger}~(HLT), is implemented in software and aims to further reduce the event rate to about 1~kHz on average.
Events passing the HLT are then stored on local disk or in CMS Tier-0\footnote{The Worldwide LHC Computing Grid (WLCG) is composed of four levels, or ``Tiers'', 
called 0, 1, 2 and 3. Each Tier is made up of several computer centres and provides a specific set of services; they process, store and analyse 
all the data from the Large Hadron Collider (LHC). Tier 0 is the CERN Data Centre. All of the data from the LHC pass through this central hub.
Tier 0 distributes the raw data and the reconstructed output to Tier 1's, and reprocesses data when the LHC is not running.}.
% The CMS High Level Trigger aims to maximize efficiency while keeping CPU-time and rate low. It must work within a peak CPU time of about 200~ms/evt at 100~kHz input rate.
% Its algorithms use the same software framework and most of the same reconstruction code used for offline reconstruction and analyses and it must be flexible to adapt to changes 
% in data-taking conditions, like changes in luminosity or special conditions occurring during the CMS commissioning or dedicated LHC fills.
% It must also provide on-line detector monitoring through specific trigger paths for calibration and alignment~\cite{nota}, and its performance should be robust with respect to 
% changes in alignment and calibration constants and stable with respect to pileup (PU).

\section{Tracking and vertexing at HLT}
\begin{figure}
\centerline{\includegraphics[width=6cm]{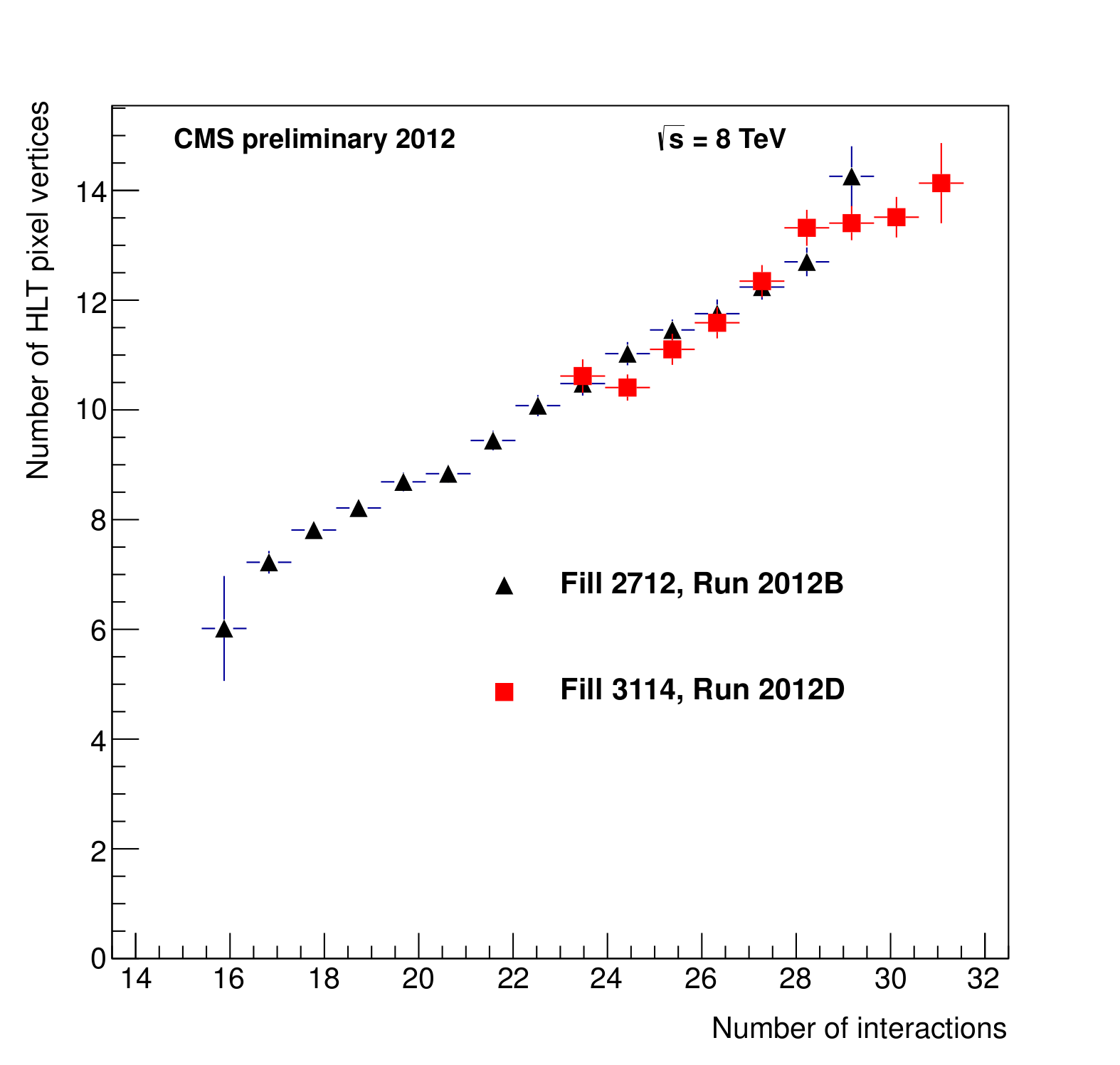}}
\vspace*{8pt}
\caption{Number of pixel vertices reconstructed at HLT.
The number of interactions is calculated from the bunch luminosity as measured by 
the forward calorimeters (HF). \label{fig:f1}}
\end{figure}

%Tracking is very important for particle reconstruction at the HLT level.
A robust and efficient tracking~\cite{Chatrchyan:2014fea} at HLT can help particle reconstruction and improve the resolution. 
For example,
%it reduces the muon trigger rate by substantially improving the momentum resolution;
%energy clusters found in the electromagnetic calorimeters can be identified as electrons or photons depending on the presence of a track; 
%the background rejection rate of the lepton triggers can be enhanced further by requiring that leptons should be isolated; 
it is possible to trigger on jets produced by b-quarks, by counting the number of tracks in a jet which have a transverse impact parameter 
incompatible with the track originating from the beam-line; or it is possible to trigger on hadronic $\tau$ decays by finding a narrow, 
isolated jet using tracks in combination with the calorimeter information.
Track reconstruction uses a significant fraction (about 30\%) of the total HLT CPU time. For this reason, track reconstruction is performed
only after other requirements have been satisfied, on about 4\% of the HLT events.
The CMS tracking software is known as the Combinatorial Track Finder (CTF)~\cite{CTF} based on the Kalman filter method~\cite{KF}. 
The collection of reconstructed tracks is produced by multiple iterations of the CTF track reconstruction sequence, in a process called \textit{Iterative Tracking}. 
In the early iterations, tracks with relatively high transverse momentum ($p_T$) and produced near the interaction region are reconstructed. After each iteration, hits associated
with tracks already found are removed, reducing the combinatorial complexity and thus allowing later iterations to search for lower $p_T$ or highly displaced tracks. 
The very first iteration is the source of most of the tracks and is designed to reconstruct prompt tracks having three pixel hits (\textit{Pixel Tracks}).\\
The reconstruction of primary vertices (PV) is also extremely important.
Pixel Tracks are used to perform a first reconstruction of the positions of the interaction points (\textit{Pixel Vertices}).
% At HLT, for the PV reconstruction a simple gap clustering algorithm is used. 
% All tracks are ordered by the $z$ coordinate of their point of closest approach to the beamspot. Then, wherever two neighbouring elements in this
% ordered set have a gap between them exceeding a distance cut $z_{sep}$, this point is used to split the tracks into two separate sets and consequently to separate vertices. 
In 2012 data taking, where up to 30 interactions per bunch crossing were registered, the number of reconstructed vertices still showed
a linear dependence on the number of interactions without saturating, as shown in Fig.~\ref{fig:f1}, where the real number of interactions is measured 
using the information from the forward calorimeter (HF).

\section{Particle Flow jets}
\begin{figure}
\centerline{\includegraphics[width=7cm]{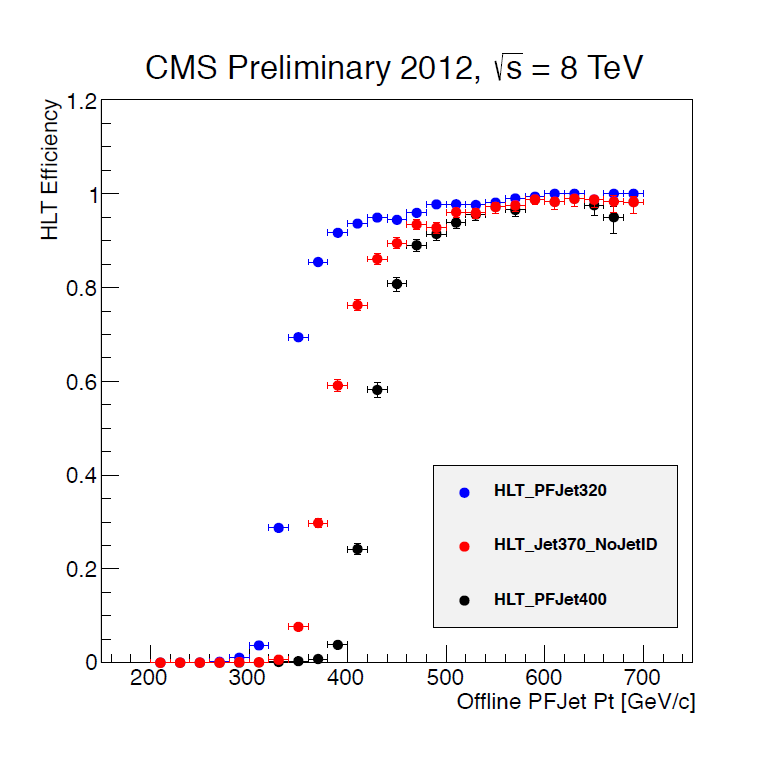}}
\vspace*{8pt}
\caption{Turn-on curve measured versus the offline Particle Flow jet $p_T$. The trigger efficiency is measured on an unbiased data sample from Run2012C. \label{f3}}
\end{figure}

At the HLT, jets are reconstructed using the anti-k$_t$ clustering algorithm with cone size $R =0.5$~\cite{antikT}. 
The input for the jet algorithm can be the calorimeter towers (called CaloJet), or the reconstructed Particle Flow objects (called PFJet). 
The Particle Flow technique allows to use the information from all the detectors and to combine them together to reconstruct the objects~\cite{PF}.
In 2012, most of the jet trigger paths used PFJets. Because of the significant CPU consumption of the Particle Flow algorithm at the HLT, PFJet
trigger paths have a pre-selection based on the CaloJet. The matching between CaloJet and PFJet is also required in single PFJet paths.
In Fig.~\ref{f3} the efficiency turn-on curve of three different trigger paths requiring PFJets with different $p_T$ thresholds are shown.

\section{$b$-tagging}
\begin{figure}
\centerline{\includegraphics[width=7cm]{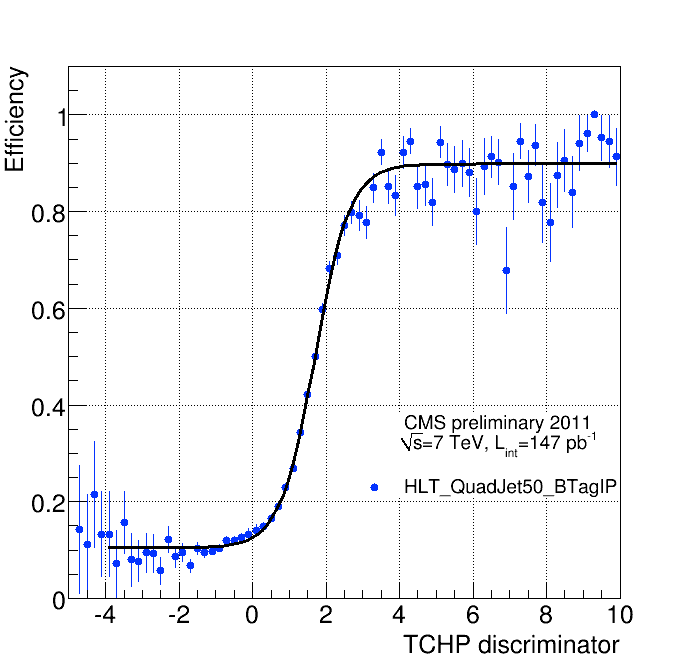}}
\vspace*{8pt}
\caption{Turn-on curves of the Track Counting High Purity (TCHP) discriminant efficiency at HLT, with respect to the same variable computed offline.\label{f4}}
\end{figure}

The precise identification of $b$-jets is crucial to reduce the large backgrounds at the LHC. In CMS, using algorithms for $b$-tagging jets~\cite{Chatrchyan:2012jua}, 
this background can already be highly suppressed at the HLT, giving lower trigger rates with large efficiency.
Algorithms for $b$-tagging exploit the fact that B hadrons typically have large decay lifetimes and the presence of leptons in the final state compared to those from 
light partons. As a consequence, tracks and vertices are largely displaced with respect to the primary vertex.
The Track Counting (TC) algorithm uses the impact parameter (IP) significance ($\sigma (IP) / IP$) of the tracks in the jets as
a discriminant to distinguish $b$-jets from other flavours. 
In Fig.~\ref{f4} the turn-on curves for the Track Counting discriminant with a High Purity requirement is shown; the online cut for this path is at TCHP$=2$.
The discriminant is defined as the third highest impact parameter significance for the tracks associated to a jet.

% %%%%%%%%%%%%%%%%%%%%%%%%%%%%%%%%%%%%%%%%%%%%%%%%%%%%%%%%%%%%%%%%%%%%%%%%%
% %%
% %%   use this format to include a LaTeX table  into your paper
% %%
% \begin{table}[t]
% \begin{center}
% \begin{tabular}{l|ccc}  
% Patient &  Initial level($\mu$g/cc) &  w. Magnet &  
% w. Magnet and Sound \\ \hline
%  Guglielmo B.  &   0.12     &     0.10      &     0.001  \\
%  Ferrando di N. &  0.15     &     0.11      &  $< 0.0005$ \\ \hline
% \end{tabular}
% \caption{ place the caption here }
% \label{tab:table1}
% \end{center}
% \end{table}
% %%%%%%%%%%%%%%%%%%%%%%%%%%%%%%%%%%%%%%%%%%%%%%%%%%%%%%%%%%%%%%%%%%%%%%%%%%%

\section{Tracking and vertexing for b-tagging in Run II}
\begin{figure}
\centerline{\includegraphics[width=8cm]{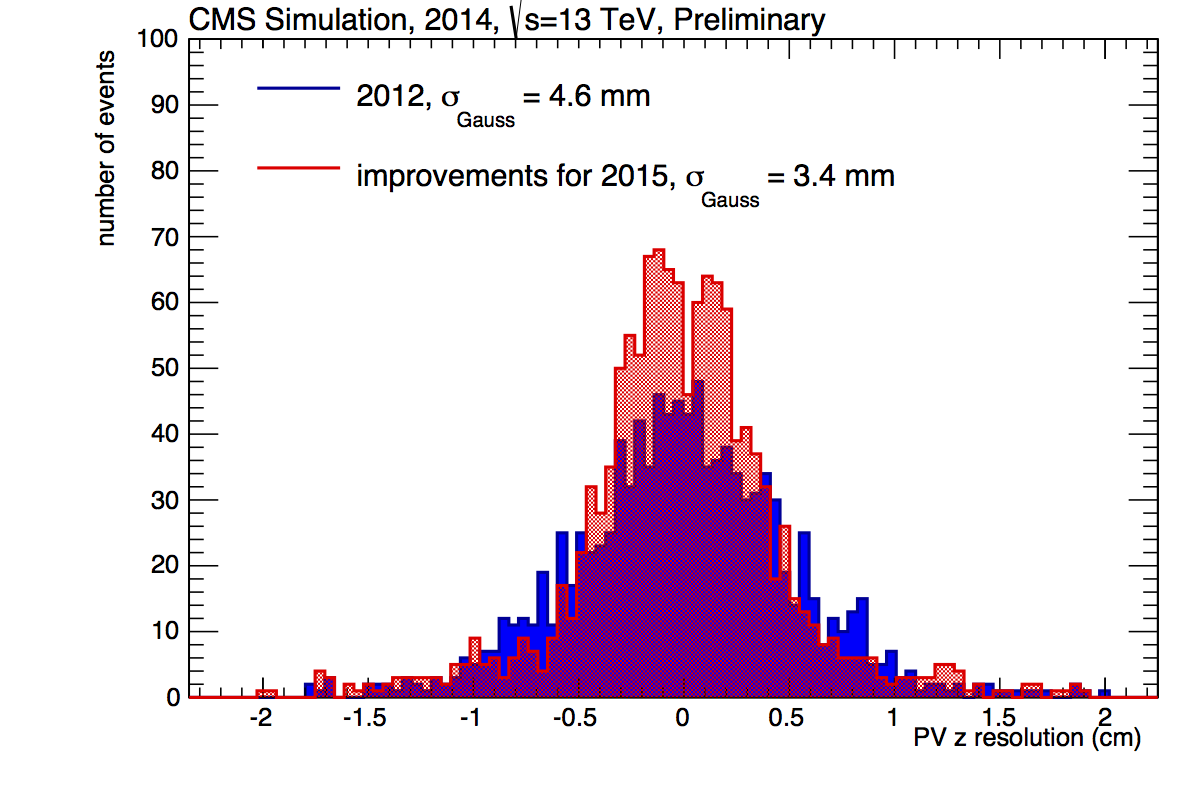}}
\vspace*{8pt}
\caption{$z$ resolution of the vertex reconstructed using FastPV, with respect to the simulated primary vertex.\label{FastPVres}}
\end{figure}

\begin{figure}[h]
\centerline{\includegraphics[width=8cm]{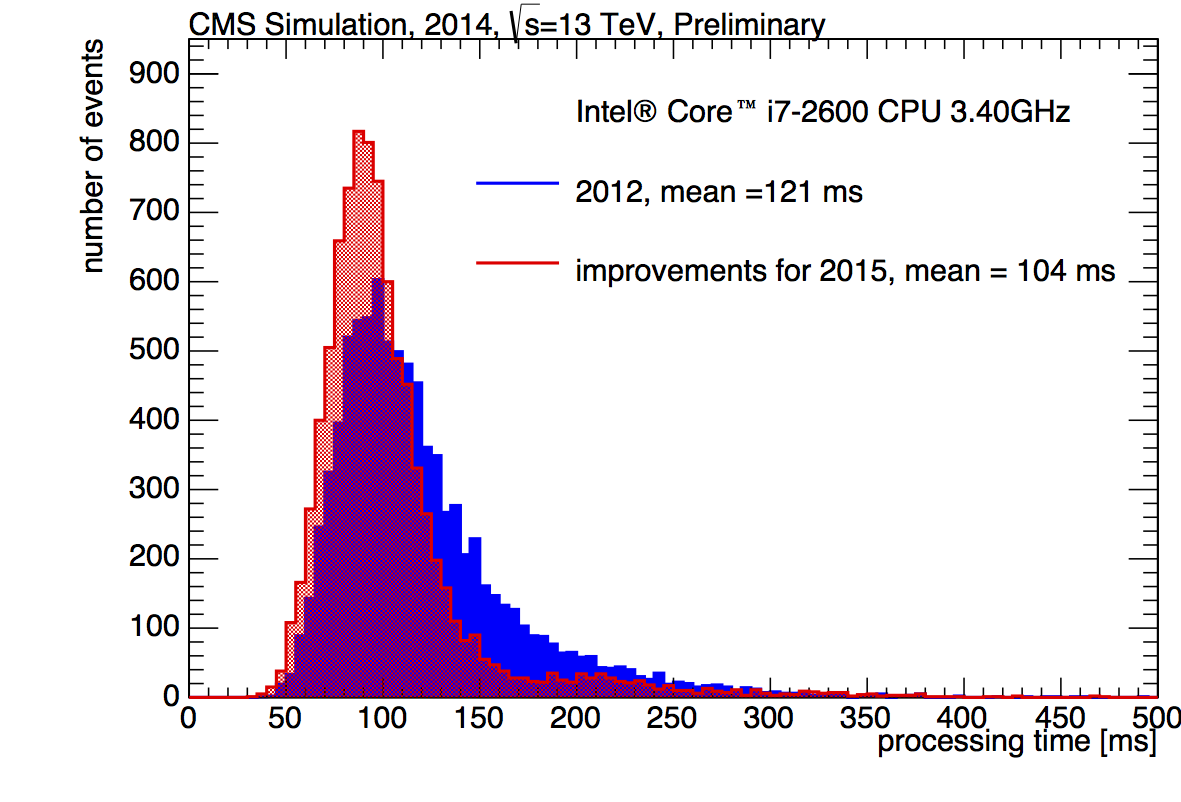}}
\vspace*{8pt}
\caption{Timing of the trigger path HLT\_DiCentralPFJet30\_PFMet80\_BTagCSV07. Two different configurations for the FastPV algorithm are compared.\label{FastPVtiming}}
\end{figure}

The Fast pixel Primary Vertex (FastPV) is an algorithm used to find the PV position before performing the reconstruction of Pixel Tracks. 
It allows to have a first fast reconstruction of the PV, in order to have a constraint for the subsequent reconstruction of the Pixel Tracks in the Iterative Tracking procedure. 
This reduces the combinatorics and aims to reach better performance.
Given a jet (with $p_T>40$~GeV) the compatible pixel clusters are selected along the jet direction. 
These clusters are projected along the jet direction onto the z axis. 
The FastPV algorithm reconstructs the PV position from the $z$ position of the highest peak. 
Once the PV position is found, full pixel tracking is performed and a pixel primary vertex is reconstructed.
%The PV reconstructed with FastPV, together with the tracks information, is used as input for b-tagging.
In Fig.~\ref{FastPVres} the resolution along the $z$ axis of the vertex reconstructed using the FastPV algorithm, with respect to the
simulated primary vertex, is shown. Simulated $Z(\nu \nu)H(bb)$ events with $<$PU$>=$ 60 are used.
The performance of the algorithm used in 2012 is compared to the improved algorithm, using weighted clusters and an extended jet acceptance ($|\eta|<2.4$ instead of $|\eta|<1$).
In Fig.~\ref{FastPVtiming} the timing of the trigger path HLT\_DiCentralPFJet30\_PFMet80\_BTagCSV07 is shown\footnote{TTbar simulated events ($\sqrt{s} =13$ TeV, $<$PU$>=$20, 
bunch spacing BS $=$ 25~ns) have been used.}.
In the configuration used in 2012, tracking was not performed according to the Iterative Tracking procedure described above, but using just a single iteration.
Primary vertices reconstructed using FastPV and tracks are used as input to b-tagging.
In the 2015 configuration, instead, the Iterative Tracking procedure is used. 

\end{document}